\documentclass[aps,prb,twocolumn,notitlepage,superscriptaddress]{revtex4-1}
\usepackage[a4paper,top=2.50cm,bottom=2.50cm,left=2.50cm,right=2.50cm]{geometry}
\usepackage{mathptmx}
\usepackage{amsmath}
\usepackage{amsfonts}
\usepackage{amssymb}
\usepackage{bm}
\usepackage{enumerate}
\usepackage{graphicx}
\usepackage{color}
\DeclareGraphicsExtensions{.eps, .pdf, .jpg, .tif}
\usepackage{hyperref}

\def\vr{{\bf r}}
\def\vv{{\bf v}}
\def\vp{{\bf p}}
\def\vg{{\bf g}}
\def\cD{{\cal D}}
\newcommand{\grad}{\pmb\nabla}
\def\text#1{\mbox{\tiny #1}}
\newcommand{\nimp}{n_{\text{imp}}}
\newcommand{\sigmatr}{\sigma_{\text{tr}}}
\newcommand{\tcoh}{t_{\text{coh}}}
\newcommand{\elltr}{\ell}
\newcommand{\swave}{$^1${\sf S}$_0$}

\newcommand{\kb}{k_{\text{B}}}
\def\nicefrac#1#2{\genfrac{}{}{}{1}{#1}{#2}}
\begin{document}
\title{Superfluidity in Disordered Neutron Stars Crusts}
\author{J. A. Sauls}
\affiliation{Center for Applied Physics \& Superconducting Technologies, 
             Department of Physics \& Astronomy,
             Northwestern University, Evanston, Illinois 60208, USA \\
and Fermi National Accelerator Laboratory, Batavia, Illinois 60510-5011, USA
}
\email{sauls@northwestern.edu}
\author{N. Chamel}
\affiliation{Institut d’Astronomie et d’Astrophysique, CP-226, 
	     Universite Libre de Bruxelles, Brussels, Belgium}
\author{M. A. Alpar}
\affiliation{Sabancı University, Faculty of Engineering \& Natural Sciences, 
             Orhanlı, 34956 Istanbul, Turkey}
\date{\today}
\begin{abstract}
Nonequilibrium conditions imposed by neutrino cooling through the liquid-solid transition lead to disorder in the solid crust of neutron stars. Disorder reduces the superfluid fraction, $\rho_s/\rho$, at densities above that of neutron drip, $\rho_d \approx 4\times 10^{11}\,g/cm^3$. 
For an amorphous solid crust the suppression of $\rho_s$ is small, except in the highest density regions of the crust.
In contrast to the strong reduction in neutron conduction predicted for coherent Bragg scattering in a crystalline crust, the disordered solid crust supports sufficient neutron superfluid density to account for pulsar glitches.
\end{abstract}

\maketitle
\noindent{\it Introduction --} 
In their article ``Towards a metallurgy of neutron star crusts'', Kobyakov and Pethick argue that the structure of neutron star crusts, conventionally assumed to be a body-centered cubic (bcc) crystal, requires revision.\cite{kob14} 
Their analysis was based on the role of interactions between the interstitial neutrons and nuclei that lead to more anisotropic crystalline phases.
Here we consider the metallurgical state of neutron star crusts based on the model of an amorphous metallic alloy. 

The effects of impurity disorder on electrical and heat transport by the relativistic electron fluid embedded in the crust has been discussed by P. B. Jones who showed that the inner crust is likely to be amorphous.~\cite{jon99} The reduced electrical conductivity of an amorphous metallic crust has important implications for the nature of the magnetic field in pulsars.~\cite{jon04a,jon04b}
Here we address the impact of impurity and structural disorder on neutron superfluidity in the crust. 

The outer crust of a neutron star contains nuclei embedded in a relativisitic electron liquid. With increasing density electron capture leads to nuclei becoming ever more neutron rich. Above the density of ``neutron drip'', $\rho_d \approx 4.3\times 10^{11}\,\mbox{g/cm}^3$, the inner crust consists of unbound neutrons, which become superfluid at temperatures below $T_c\approx 10^9\,\mbox{K}$, in addition to relativistic electrons embedded in a nuclear solid. 
At a density of order $\rho_c\approx 2\times 10^{14}\,\mbox{g/cm}^3$, the crust dissolves into a mixture of superfluid neutrons, relativistic electrons, and superconducting protons.\cite{bay75a}

Quantized vortices in the neutron superfluid and quantized flux lines in the proton superconductor are topologically stable defects of these hadronic fluids.\cite{sau89} Their evolution under non-equilibrium conditions imposed by neutrino cooling, rotation and spin-down of the solid crust determines the response of the superfluid components of neutron stars to the radiative spin-down, and to pulsar glitches.\cite{alp77,and82,alp84a,alp84,alp88,rud98,has15,piz16,wat17,gra18}

Interactions between nuclei and the neutron superfluid vortices in the inner crust are thought to be central to establishing the critical-state responsible for pulsar glitches.\cite{alp77,alp81a} Glitch observations are informative probes of neutron star structure.\cite{alp84a}
The spatial organization of nuclei in the neutron star crust, and their interactions with the superfluid neutrons and quantized vortices determine the transport properties relevant to the rotational dynamics of neutron stars.\cite{pin85}
The relatively small offset in the postglitch spin-down rate, $\Delta\dot\Omega_c/\dot\Omega_c\approx I_s/I\lesssim10^{-2}$ for many pulsars, is consistent with most of the superfluid - the core superfluid - being coupled to the crust and plasma on timescales much shorter than the observed limits on the glitch rise time and the postglitch recovery time-scales.~\cite{alp84}
Thus, the neutron superfluid interacting with nuclei in the inner crust has been identified as the component in the star comprising the right fraction, $I_s/I$, of the moment of inertia that is responsible for both glitches and postglitch relaxation.\cite{alp84a}

The neutron superfluid fraction in the crust, $n_s/n$, has received renewed attention as a result of band-structure calculations for the interstitial neutrons in the coupled system of neutron liquid and nuclei at densities above that of neutron drip.\cite{cha05,cha06,cha12}
These calculations have been interpreted as a suppression of the crust neutron superfluid moment of inertia compared to pure neutron matter by an order of magnitude due to coherent Bragg scattering by a crystalline lattice of nuclei.\cite{cha12} 
Such a large reduction in the superfluid density necessitates a revision of the crust superfluid as the sole source of pulsar glitches and the slow postglitch relaxation.\cite{and12,cha13} See also Refs.~\onlinecite{noe16,wat17}.

However, the non-equilibrium conditions under which the crust forms in a hot young neutron star may prevent formation of a crystalline solid.
In this Letter we address the impact of a disordered solid crust on neutron superfluidity based on a ``metallurgy of neutron star crusts'' adapted from the theory of ``metallic alloys'', ``amorphous metals'' and ``dirty superconductors''.
The central questions we address are ``how do quenched impurities and structural disorder in the solid nuclear crust impact our understanding of superfluidity of the interstitial neutrons in the inner crust?'' 
and, ``what is the fraction of the neutron superfluid in the inner crust that contributes to the slow postglitch relaxation?".
The main conclusion of our analysis is the superfluid fraction of the crust moment of inertia, even in an amorphous crust, is sufficient to account for observed glitch signatures without invoking a de-coupled superfluid component within the core of the star.

\smallskip

\noindent{\it Quenched Solidification --} 
Neutron stars are born ``hot'', with interior temperatures of order $10^{12}\,\mbox{K}$. They cool rapidly via neutrino emission processes. Even for relatively slow cooling by the modified URCA process the neutrino emissivity scales as $\varepsilon_{\mbox{\small mURCA}}=10^{21}\,T_9^8\,\mbox{erg}/\mbox{cm}^3\mbox{-s}$. At $T\approx 10^{11}\,\mbox{K}$ this rate is comparable to the neutrino luminosity from direct URCA processes.\cite{pag06}
Thus, at temperatures of order the onset of crystallization in the crust, $T_{\text{X}}\approx 10^{11}\,\mbox{K}$,~\footnote{The onset of crystallization (local order) for a Coulomb lattice is discussed in Shapiro and Teukolsky, c.f. text and Eq. 4.3.21 on page 91 of Ref.~\onlinecite{shapiro83}. For $Z=40$ and crust densities above neutron drip $T_{\text{X}}$ spans the range $1\times 10^{11}\,-\, 7\times 10^{11}\,\mbox{K}$.} a cubic centimeter of nuclear matter is radiating energy via neutrinos at a rate of $\dot{Q}_{\nu}\approx 10^{37}\,\mbox{erg}/\mbox{s}$.

The specific heat of degenerate neutron matter at density $\rho\approx2\times 10^{14}\,\mbox{g}/\mbox{cm}^3$ is $\bar{C}_v = 1.5\times 10^{20}\,T_9\,\mbox{erg}/\mbox{K}\mbox{-cm}^3$. Thus, in this density range of the crust the cooling rate of a cubic centimeter of neutron star matter is $\dot{T}=-\dot{Q}_{\nu}/\bar{C}_v\approx -6.7\times T_9^7\,\mbox{K}/\mbox{s}$ $\approx -6.7\times 10^{14}\,\mbox{K}/\mbox{s}$ at $T=10^{11}\,\mbox{K}$, corresponding to a quench rate at the onset of crystallization of $1/\tau_{Q_{\text{X}}}= -\dot{T}\vert_{T_{\text{X}}}/T_{\text{X}} \approx 6.7\times 10^{3}\,\mbox{s}^{-1}
\,\mbox{at}\,T_{\text{X}} = 10^{11}\,\mbox{K}$.
The temperature drops from $T_{\text{X}}$ to $T_{\text{X}}/2$ in a milli-second. Density variations in the inner crust lead to inhomogeneity of cooling and quench rates. 
Thus, the conditions present at the onset of solidication in the inner crust are far from the idealized conditions necessary for growth of a crystalline crust; rather the crust is likely disordered.\cite{kle60,ino00}

In addition to structural disorder, impurity disorder in which nuclei with neutron-proton ratios that are out of beta equilibrium are also important. Quenched fluctuations in proton number, $Z$, at fixed nucleon number, $A$, are expected to occur in the early history of the neutron star cooldown.~\cite{jon99} 
Proton pairing effects from strong nuclear forces lead to extremely small formation enthalpies for a range of proton number deviations, $\delta Z=\pm 2,\pm 4\,\ldots$.~\cite{jon99} 
Calculations of the difference in Gibbs energy between nuclei that differ in proton number by $\delta Z$ are of order $\delta G_{\mbox{\small bind}} \approx 0.01 - 0.1 \mbox{MeV}$ per nucleus.
Thus, even for equilibrium nuclear matter the relative population of an impurity to ground-state nucleus, $N(A,Z-\delta Z)/N(A,Z)=\exp(-\delta G_{\mbox{\small bind}}/k_B T)$, is near equality at temperatures of order $T_{\text{X}}$; even at temperatures below the superfluid transition temperature in the inner crust, $T\approx 10^9\,\mbox{K}$, the equilibrium population of charge impurities is significant.

\medskip

\noindent{\it Superfluidity in the Crust --}
Superfluidity of interstitial neutrons by binding and condensation of s-wave, spin-singlet ($^1S_0$) Cooper pairs onsets at temperatures, $T_c\approx 10^9 - 10^{10}\,\mbox{K}$, with an order parameter of the form $\Psi(\vr)=\langle \psi_{\uparrow}(\vr)\psi_{\downarrow}(\vr)\rangle=|\Psi(\vr)|\,e^{i\vartheta(\vr)}$, where $\psi_{\uparrow}(\vr)\psi_{\downarrow}(\vr)$ is the operator for $^1S_0$ Cooper pairs.\cite{sau89}
For spatial variations on lengthscales longer than the radial extent of a Cooper pair, $\xi_0 = \hbar v_f /\pi\Delta$ where $2\Delta$, is the binding energy of a Cooper pair, the condensate density is to good approximation given by the ground-state value, $|\Psi|^2\sim n$. 
The corresponding superfluid mass current is the response to a gradient of the phase of the neutron condensate order parameter. To linear order in the phase gradient the superfluid mass current is $\vg_s = n_{s} \vp_s$, where $\vp_s=\nicefrac{\hbar}{2}\grad\vartheta$ is the neutron condensate momentum per particle, and $n_{s}$ is the neutron ``superfluid density'', i.e. the dissipationless component of the momentum density of the neutron fluid. 
The superfluid density, $n_s$, is a response function, and in general is not equivalent to the static equilibrium condensate fraction $|\Psi|^2_{eq}$. Excitations out of the condensate contribute to the ``normal'' neutron component of the current. Thus, the hydrodynamics of the neutron superfluid requires a second, generally dissipative, ``normal'' component with density $n_n$ and momentum per particle, $\vp_n$, such that the total mass current density is $\vg = n_s\vp_s + n_n\vp_n$.~\cite{london54} 

For pure neutron matter the opening of a gap, $\Delta$, in the excitation spectrum freezes out the neutron quasiparticle excitations at temperatures $T\ll\Delta$ with $n_n\propto e^{-\Delta/\kb T}$. In the zero temperature limit the normal component of a fully gapped, translationally invariant superfluid vanishes. Thus, the superfluid density in the zero temperature limit is the full neutron density. This is a special case in which Galilean invariance ensures $\lim_{T\rightarrow 0}n_s=n$.\footnote{This result is also based on the assumption of linear response for the mass current. In particular, the dependence of the supercurrent on the gap reappears for sufficiently large phase gradients. Indeed the supercurrent collapses catastrophically at the critical momentum $|\vp_s| \ge p_c = \Delta/v_f$.} Galilean invariance also allows us to introduce the superfluid velocity field, $\vv_s = \nicefrac{\hbar}{2 m_n}\grad\vartheta$, where $m_n$ is the bare neutron mass; $\vv_s$ transforms as a velocity field under a Galilean boost. For pure neutron matter in the rest frame of the excitations the mass current can be expressed as $\vg = \rho_s\,\vv_s$ where $\lim_{T\rightarrow 0}\rho_s = \rho$ is the neutron mass density.

In the neutron star crust the magnitude of the mass current of neutrons moving \emph{relative} to the nuclear solid is not protected by Galilean invariance. The presence of nuclei embedded in the neutron liquid breaks Galilean invariance. The interaction between the interstitial neutrons and the crustal nuclei leads to scattering, pair-breaking, and to reduction of the superfluid mass fraction, $\rho_s/\rho < 1$, even at $T=0$.

\medskip

\noindent{\it Crustal Alloy Model --} 
What are the effects of a random nuclear potential on superfluidity in the inner crust? Quantitative answers depend on the nature of the disorder, but much can be understood in the context of statistical models of \emph{uncorrelated random impurities} that have been widely studied in the context of superconductivity in metallic alloys.~\cite{edw58,and59,abr59a,abr59b}

Consider first the situation in which a fraction of the nuclei are impurities. The crust is a ``nuclear alloy'' in which two or more nuclear species are embedded at random positions within the neutron liquid, a situation analogous to terrestrial metallic alloys such as Cu-Zn.
Itinerant neutrons experience a random nuclear potential characterized by the density of nuclear impurities, $\nimp$, and the differential cross-section, $d\sigma/d\Omega_{\vp'}(\vp',\vp)$, 
for neutrons of momentum $\vp$ scattering elastically off a nuclear impurity into the final state $\vp'$.

Disorder impacts neutron transport in the normal state even at temperatures well below the neutron Fermi temperature, $T_f=E_f/\kb\approx 10^{11}\,\mbox{K}$. The key physical parameter is the neutron transport mean free path (\emph{mfp}), $\elltr = 1/\nimp\sigmatr$, where $\sigmatr$ is the transport cross-section for elastic scattering of neutron quasiparticles off impurity nuclei, $\sigmatr=\int d\Omega_{\vp'} d\sigma/d\Omega_{\vp'}(\vp',\vp)(1-\hat\vp\cdot\hat\vp')$. Nuclear radii are in the range $R_N\approx 6 - 7 \,\mbox{fm}$. For neutron-nucleus scattering we are generally in the limit $k_f R_N \gg 1$ in which case the transport cross-section is to good approximation given by $\sigmatr = \pi R_N^2$. The corresponding transport mean-free path is determined by the density of impurity scattering sites. In particular low-temperature normal-state mass transport is governed by the diffusion equation with diffusion constant, $\cD = \nicefrac{1}{3}\,v_f\,\elltr$. Low temperature heat transport by neutrons is limited by the same scattering process, with the thermal conductivity of interstitial neutron matter given by $\kappa_n = \cD\,C_v = \nicefrac{1}{3}\,v_f\,C_v\,\elltr$. Nuclei in the crust are relatively far apart in which case the measure of disorder for the normal Fermi liquid remains a small parameter, i.e. $1/k_f\elltr \ll 1$, throughout the inner crust.

For the superfluid state a new length and time scale appears. The timescale for Cooper pair formation from the normal Fermi-liquid ground state is $\tcoh=\hbar/\pi\Delta(0)$, where $\Delta(0)\simeq 1.78\kb T_c$ is the zero-temperature BCS energy gap. The corresponding spatial coherence length for the pairing correlations is given by $\xi_0 = v_f\,\tcoh$ for ballistic neutrons moving with the Fermi velocity. The pairing correlation length scale, $\xi_0\approx 10-300\,\mbox{fm}$ through the inner crust, is typically much longer than the Fermi wavelength. 

In a disordered alloy the ballistic pair correlation length is invalid for $\elltr\lesssim\xi_0$. There is a ``dirty limit'' in which the correlation length is strongly reduced because of the diffusive motion of the neutron quasiparticles that form Cooper pairs, i.e. $\xi^2=3\cD\tcoh$. The pairing correlation length is then given by $\xi=\xi_0\sqrt{\elltr/\xi_0}$. The key observation is that the scale for disorder of the nuclei to impact the superfluid state is set by $\elltr/\xi_0$, the ratio of the transport mean free path to the clean limit correlation length. 
In the strong disorder limit, $\elltr\ll\xi_0$, the superfluid density is dramatically suppressed by pair-breaking effects of scattering in the presence of condensate flow, $n_s\rightarrow n\,\elltr/\xi_0$. However, as we show below, this strong disorder result for $n_s/n$ is only realized in high-density regions of the inner crust. The extension to any value of the disorder parameter, $\alpha=\nicefrac{\pi}{2}\xi_0/\elltr$, is 
\begin{eqnarray}\label{eq-ns_vs_alpha}
n_s/n 
&=& \int_{0}^{\infty} dE\,
\frac{\Delta^2}
     {(E^2+\Delta^2)\left(\sqrt{E^2+\Delta^2}+\alpha\,\Delta\right)}
\nonumber\\
&\approx&
\left\{
\begin{matrix} 
\displaystyle{1 - \frac{\pi^2}{8}\frac{\xi_0}{\elltr}}\,, & \alpha \ll 1 \,,
\\
\displaystyle{\frac{\elltr}{\xi_0}}\,,\qquad              & \alpha \gg 1 \,.
\end{matrix}
\right.
\end{eqnarray}
This result, and the theory of disorder effects on the crust superfluid based on Eilenberger's formulation of Gorkov's equations for disordered superconductors, is discussed in Ref.~\onlinecite{sau20c}.

To make predictions for the suppression of the superfluid fraction throughout the inner crust we need a theoretical model for the density and cross section for scattering of neutron quasiparticles, in addition to the \swave\ neutron superfluid gap and coherence length. The gap, coherence length, mean distance between nuclei and radii of the nuclei as a function of density are provided in the
Appendix.
%
We present results for the neutron superfluid embedded in an \emph{amorphous solid crust}, analogous to metallic glasses, strongly disordered metals created by thermal quench.\cite{kle60,ino00} Disorder is then defined on the scale of the mean inter-atomic distance.

\medskip

\noindent{\it Amorphous Crust Model --} 
Thermal quench by neutrino cooling, combined with density gradients and long timescales for diffusion in the crust favor a non-crystalline crust, and a disordered nuclear solid crust embedded in a degenerate neutron fluid. There is no crystalline order and the neutron/proton ratios in nuclei are also out of equilibrium with respect to both density and temperature. 

\begin{figure}[t]
\includegraphics[width=\columnwidth]{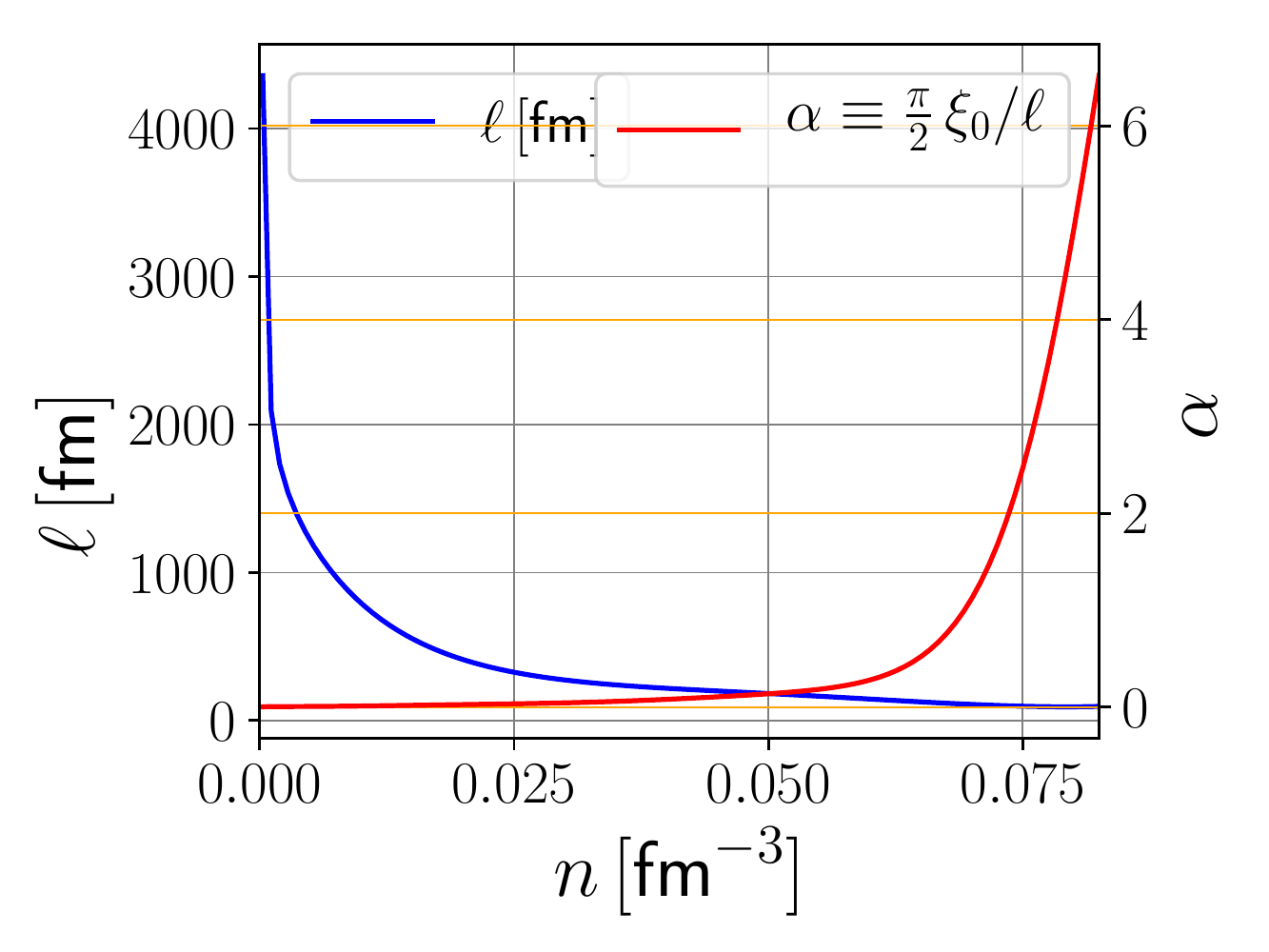}
\caption{The transport \emph{mfp}, $\elltr$, for neutrons in an amorphous nuclear solid crust 
is shown with the blue line as a function of neutron density. The \emph{mfp} varies from $\ell_{\mbox{\tiny max}}=4360\,\mbox{fm}$ at low density to $\ell_{\mbox{\tiny min}}=80\,\mbox{fm}$ at high density.
The corresponding pair-breaking parameter, $\alpha=\nicefrac{\pi}{2}\xi_0/\elltr$, 
is shown as the red line.
\label{fig-mfp+alpha_vs_n}}
\end{figure}

The amorphous nuclear crust is assumed to have the same average density of nuclei as that of a crystalline crust. Thus, the mean density of nuclei is the inverse of the Wigner-Seitz cell volume $V_{\mbox{\tiny WS}} = \frac{4\pi}{3}R_{\mbox{\tiny WS}}^3$; the Wigner-Seitz radius $R_{\mbox{\tiny WS}}$ is provided in Fig.~2 of the 
Appendix.
as a function of neutron density throughout the crust.~\cite{ons08} Since each nucleus is randomly displaced from its equilibrium position the mean density of scattering centers, ``impurities'', is $n_{\mbox{\tiny imp}} = 1/V_{\mbox{\tiny WS}}$. 

\begin{figure}[t]
\includegraphics[width=\columnwidth]{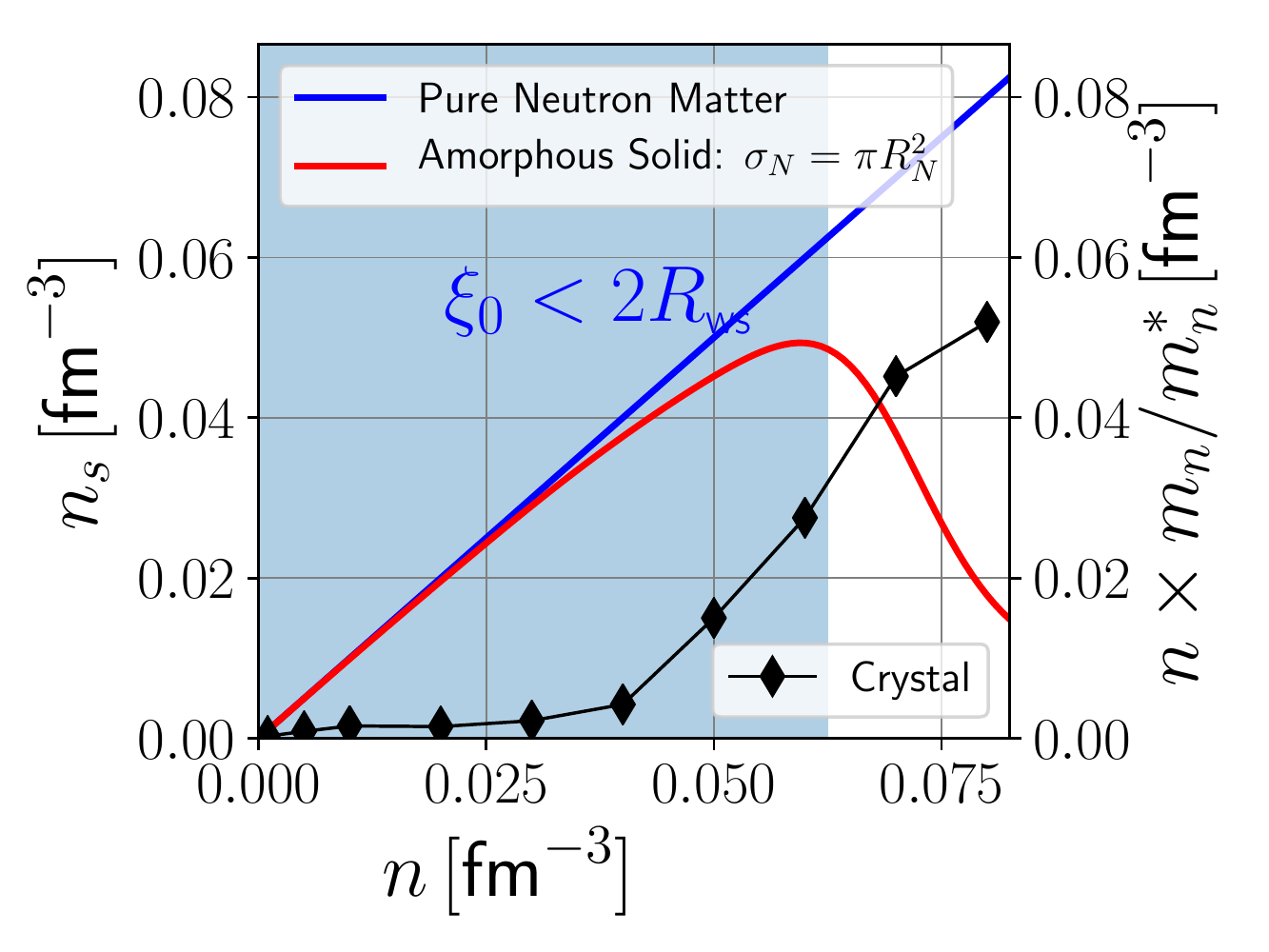}
\caption{Suppression of the zero-temperature neutron superfluid density in an amorphous crust (red line). The blue line, $n_s = n$, is for pure neutron matter at $T = 0$.
The prediction from Ref.~\onlinecite{cha05} of the conducting neutron density, $n_c=n\times m_n/m_n^*$, obtained from the band effective mass ratio for a bcc crystal of nuclei, is shown with black diamonds.
The shaded region is where the pure neutron superfluid coherence length is \emph{less} than the distance between nuclei. 
\label{fig-ns_vs_density-glass_model}}
\end{figure}

In contrast to the strong density dependence of the inter-nuclear distance, the radii of nuclei throughout the crust are weakly dependent on density, in the range $R_{\mbox{\tiny N}}\approx 6-7\,\mbox{fm}$ from low- to high-densities within the inner crust.
The transport mean free path is defined by $\elltr = 1/n_{\mbox{\tiny imp}}\,\sigma_{\mbox{\tiny tr}} = 
\nicefrac{3}{4}\,R_{\mbox{\tiny WS}}\left(R_{\mbox{\tiny WS}}/R_{\mbox{\tiny N}}\right)^2$.
The mean-free-path for an amorphous nuclear crust, and the pair-breaking parameter $\alpha=\nicefrac{\pi}{2}\xi_0/\elltr$, are shown in Fig.~\ref{fig-mfp+alpha_vs_n}.
The effect of disorder is weak ($\alpha \ll 1$) throughout most of the crust, and most significant ($\alpha > 1$) at densities above $n\approx 0.065\,\mbox{fm}^{-3}$. The corresponding effect on the neutron superfluid density is shown in Fig.~\ref{fig-ns_vs_density-glass_model}. There is only weak suppression of the superfluid density relative to the free neutron density in the weak pair-breaking limit ($n \lesssim 0.06\,\mbox{fm}^{-3}$), but at higher densities we cross-over to the ``dirty limit'' with $\alpha > 1$ and the superfluid fraction, $n_s/n$ is strongly suppressed, with a maximum suppression of the superfluid density from $n_s^{\mbox{\tiny pure}}\approx 0.081\,\mbox{fm}^{-3}$ to $n_s^{\mbox{\tiny dirty}}\approx 0.015\,\mbox{fm}^{-3}$. 

It is striking that in this strong disorder limit the impact of the disorder is relatively weak 
except at the highest densities of the inner crust. The reason is that in the region of the inner crust with $n\lesssim 0.06\,\mbox{fm}^{-3}$ nuclei are well separated, $d\approx 2R_{WS}\gtrsim 40\,\mbox{fm}$, and their cross-sections for neutron scattering are set by their much smaller radii, $R_{\mbox{\tiny N}}\approx 6\,\mbox{fm}$. This is also the region of relatively short coherence length Cooper pairs, $\xi_0\approx 10-15\,\mbox{fm}$ (shaded region of Fig.~\ref{fig-ns_vs_density-glass_model}).
Thus, even though the crust is disordered on the scale of the mean inter-nuclear distance, $\elltr\gtrsim d$, we are in the weak pair-breaking limit ($\alpha \lesssim 1$) for superfluid transport, except at the highest densities of the inner crust.
These features - neutron superfluidity and crust metallurgy - differentiate neutron star crusts from nearly all terrestrial superconducting alloys. In the latter case the interatomic distances between nuclei are generally much smaller than the superconducting coherence length, in which case superconductivity in amorphous metals is in the dirty limit $\elltr\ll\xi_0$.

Another peculiarity of the neutron star crusts related to the large internuclear distance is the result for the ``conducting neutron density'', $n\times m_n/m_n^*$ (black diamonds in Fig.~\ref{fig-ns_vs_density-glass_model}), obtained from band-structure calculations for neutrons scattering off a perfectly crystalline BCC lattice of nuclei.~\cite{cha12}
This result has been interpreted to imply a substantial reduction in the superfluid fraction of neutrons, which led to the inference that the crust does not contain enough superfluid to account for pulsar glitch obervations. It was proposed that some fraction of the \emph{core superfluid}, perhaps confined by toroidally aligned magnetic flux tubes, must be responsible for pulsar glitch responses.\cite{cha13,gug14}

An order of magnitude suppression of $n_s/n$ for a crystalline crust in regions where $\xi_0 < d$ and $R_{\text{N}} \ll d$ is in sharp contrast to the results for $n_s/n$ in an amorphous crust with the same nuclear density. 
The key observations are that disorder destroys coherent neutron Bragg scattering, yet remarkably, even in the limit of strong disorder on the scale of the inter-nuclear distance (amorphous crust model), most of the inner crust is in the weak pair-breaking limit with $\lim_{T\rightarrow 0}n_s/n \lesssim 1$. Only in the high-density region of the inner crust is the superfluid fraction substantially suppressed by neutron-nucleus scattering, c.f. Fig.~\ref{fig-ns_vs_density-glass_model}.

\medskip

\noindent{\it Summary \& Conclusion} -- 
Cooling by neutrino emission through the onset of solidification to temperatures well below that of the superfluid transition likely leads to impurity and structural disorder in the neutron star crust, analogous to amorphous metals and dirty superconductors.
Disorder destroys coherent Bragg scattering of neutrons by  nuclei. A disordered nuclear solid also destroys the local Galilean invariance of pure neutron matter leading to sub-gap quasiparticle formation (pair-breaking) for current carrying states and a reduction of the fraction of the neutron density that can support superflow.  
However, even for a strongly disordered, amorphous, nuclear solid the disorder is weakly pair-breaking over most of the inner crust.
The impact of our result on pulsar glitches is measured by the superfluid fraction of the momentum of inertia of the crust, $I_s/I_c$. Results for $I_s/I_c$ for pure superfluid neutron matter, neutrons entrained by Bragg scattering and superfluid neutrons embedded in an amorphous nuclear solid are provided in the 
Appendix.
%
The central conclusion of our analysis is the superfluid fraction of the crust moment of inertia in a disordered nuclear crust is {\it at most} reduced by $I_s^{\mbox{\tiny amorphous}}/I_s^{\mbox{\tiny pure neutrons}}\approx 89\,\%$ compared to that for pure neutron matter, and thus able to account for observed glitch signatures without invoking a decoupled superfluid component within the core of the star.

\medskip
\noindent{\it Acknowledgements} -- 
This work was initiated and completed at the Aspen Center for Physics, which is supported by National Science Foundation grant PHY-1607611.
The research of JAS is supported by US National Science Foundation Grant DMR-1508730, 
and 
that of NC by Fonds de la Recherche Scientfique (BE) Grants No. CDR J.0115.18 and No. PDR T.004320.

\begin{tabular}{c}
\qquad\qquad\qquad\qquad\qquad\qquad\qquad\qquad\qquad
\\
\hline
\qquad\qquad\qquad\qquad\qquad\qquad\qquad\qquad\qquad
\end{tabular}

\begin{figure}[t]
\begin{minipage}{0.45\textwidth}
\includegraphics[width=\columnwidth]{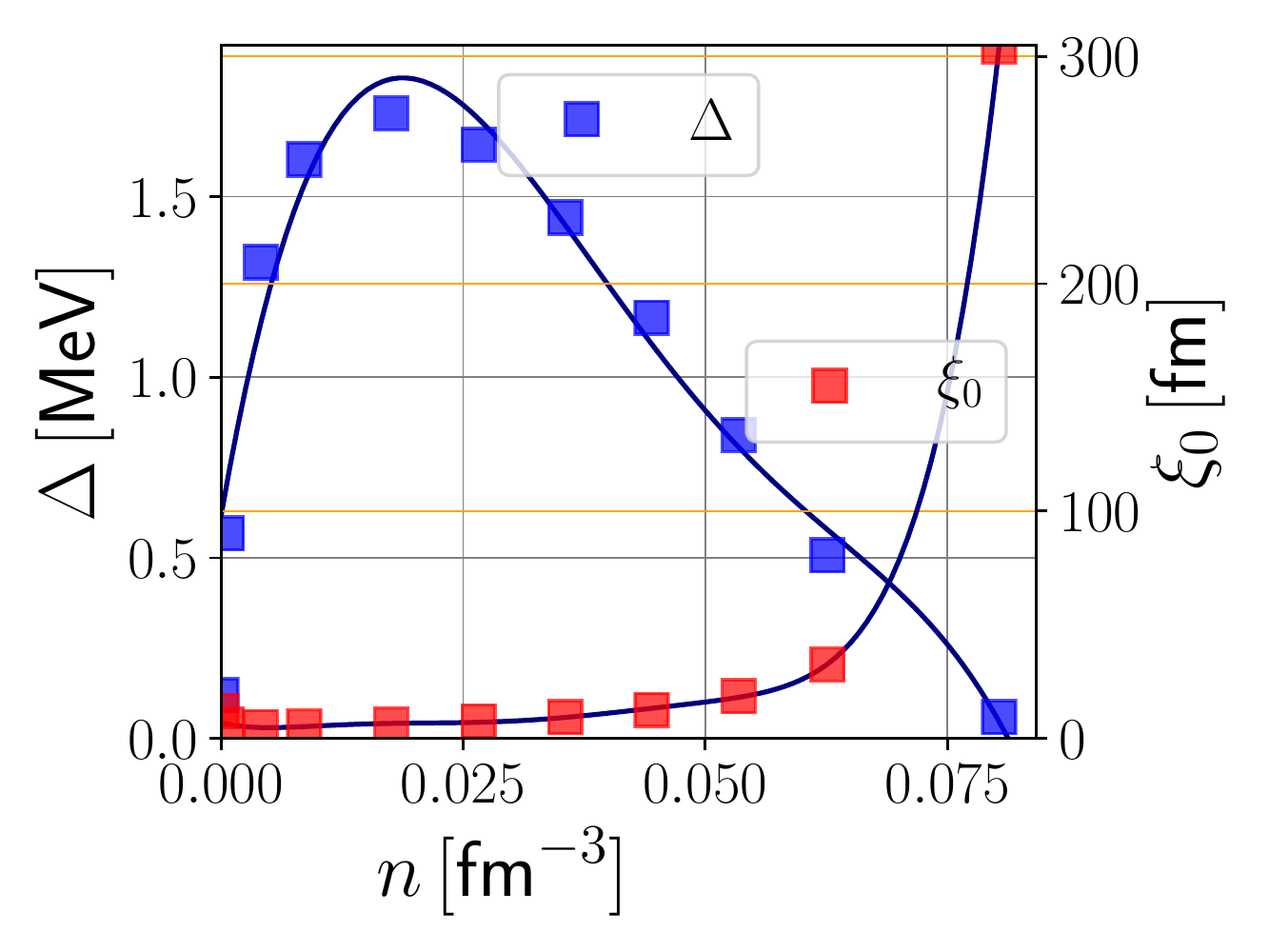}
\caption{
The zero-temperature gap amplitude, $\Delta$, and Cooper pair correlation length, $\xi_0$, for the $^1S_0$ neutron superfluid in the clean limit. Data for $\Delta$ and $v_f$ are from Ref.~\onlinecite{cao06}.
\label{fig-Delta+xi0_vs_density}
}
\end{minipage}
\end{figure}

\begin{figure}[t]
\begin{minipage}{0.45\textwidth}
\includegraphics[width=\columnwidth]{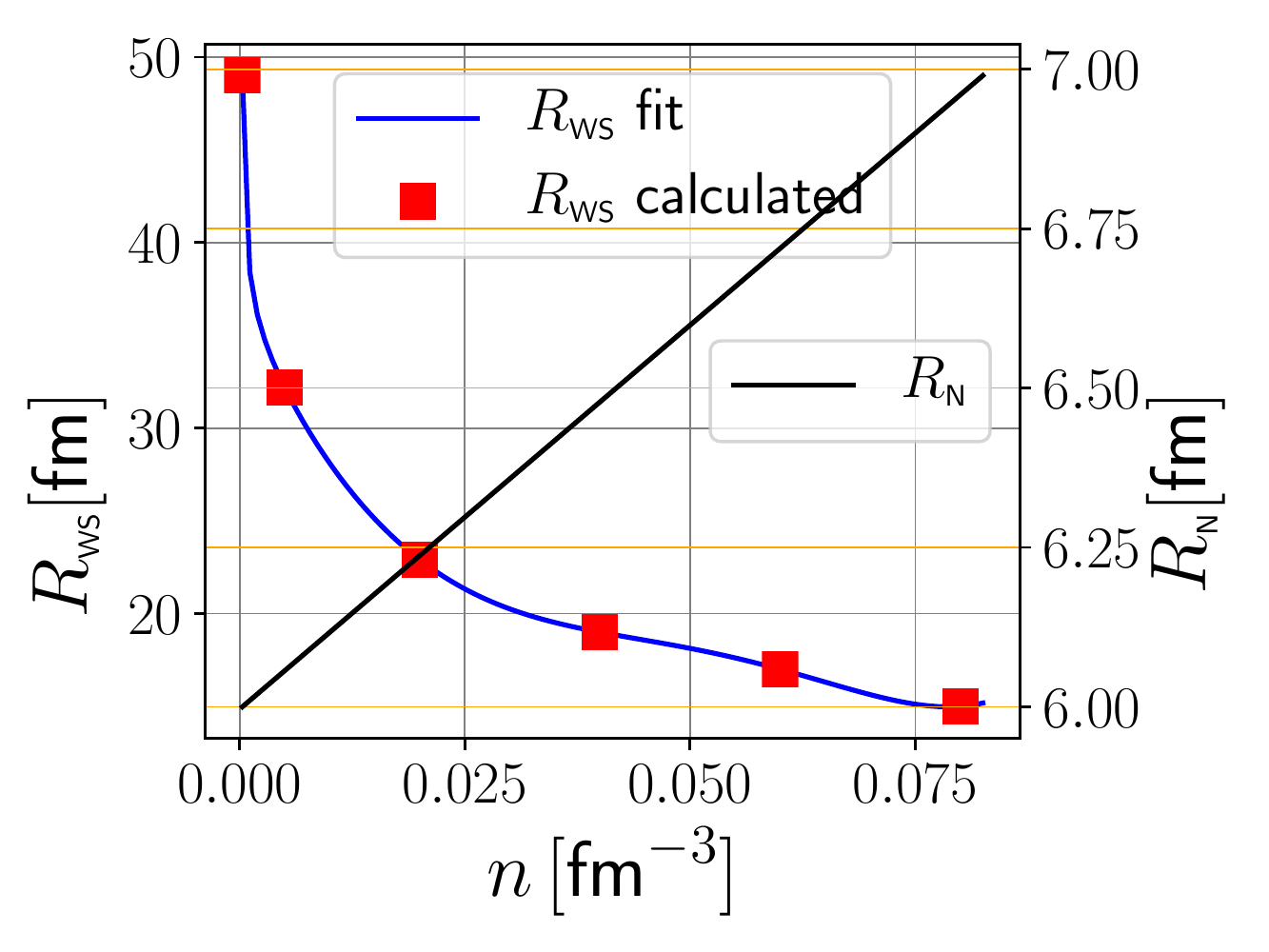}
\caption{The Wigner-Seitz radius for a crystalline nuclear lattice as a function of neutron density throughout the crust. In the nuclear glass model the mean distance between the nuclei is $d=2R_{ws}$. Data are from Ref.~\onlinecite{cha12}.
\label{fig-Rws_vs_n}}
\end{minipage}
\end{figure}

\noindent{\it Appendix: The $^1$S$_0$ Gap \& Crust Composition --}
In order to calculate the effects of disorder in the crust on the interstitial neutron superfluid fraction we need the Fermi velocity for neutron quasiparticles and the zero-temperature $^1$S$_0$ neutron gap function as a function of the unbound neutron density in the crust. 
These input parameters are obtained from the calculations reported in Ref.~\onlinecite{cao06}. The density dependence of the superfluid coherence length in the clean (ballistic) limit is given by $\xi_0 = \hbar v_f/\pi \Delta$. Both the gap and coherence length are plotted as a function of neutron density in Fig.~\ref{fig-Delta+xi0_vs_density}.
Note that the coherence length in the shallower region of crust (low neutron density) is as small as $\xi_0^{\mbox{\tiny min}}\approx 5\,\mbox{nm}$, but increases to $\xi_0^{\mbox{\tiny max}}\approx 300\,\mbox{nm}$ at high density in the deeper region of the crust. Thus, the inner crust, with weaker pairing gaps and longer correlation lengths, is the region of the crust that is more sensitive to pair-breaking and suppression of the superfluid fraction.
The mean density of nuclei and the nuclear radii as a function of density are also key inputs to determining the neutron mean free path in the amorphous crust model. This data is summarized in Fig.~\ref{fig-Rws_vs_n} using the same crust model of Ref.~\onlinecite{ons08} as applied in Ref.~\onlinecite{cha12}. A key observation is the mean distance between nuclei is always large compared to the neutron Fermi wavelength, and is even large compared to the neutron coherence length throughout most of the inner crust. The low density of nuclei is responsible for long mean free paths compared to the coherence length except in the highest density region of the inner crust.  

\noindent{\it Crust Superfluid Moment of Inertia --}\label{App-Is_vs_Pressure}
The pulsar glitch response is related to the superfluid fraction of the momentum of inertia of the crust, which 
is calculated based on Eq.~(9) of Ref.~\onlinecite{cha13},
\begin{equation}\label{eq-Is_vs_Pressure}
\frac{I_s}{I_c} = \frac{1}{P_{\mbox{\tiny crust}}} 
\int_{P_{\mbox{\tiny drip}}}^{P_{\mbox{\tiny crust}}}\,dP\,\frac{n_{s}(P)}{n_{b}(P)}
\,,
\end{equation}
where $n_{b}(P)$ [$n_{s}(P)$] is local baryon (neutron superfluid) density in the crust. 
Integration is carried out in the inner crust, from neutron drip ($P_{\mbox{\tiny drip}}$) to the crust-core boundary ($P_{\mbox{\tiny crust}}$), where the integration variable is the degeneracy pressure, $P$. 
\begin{figure}[t]
\includegraphics[width=0.9\columnwidth]{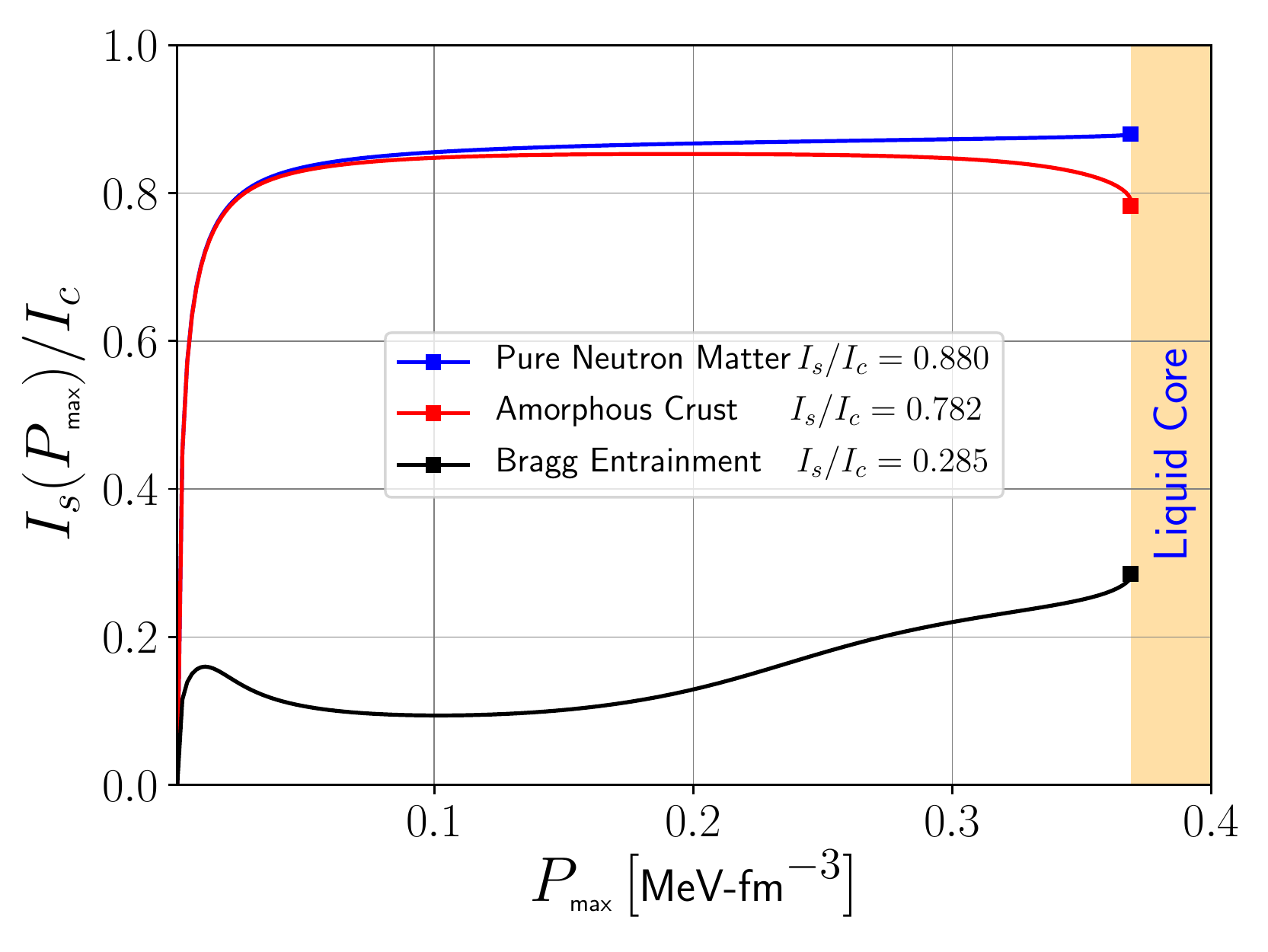}
\caption{Superfluid fraction to the crust moment of inertia as a function of $P_{\mbox{\tiny max}}$ over the range: $P_{\mbox{\tiny drip}} \le P_{\mbox{\tiny max}} \le P_{\mbox{\tiny crust}}$. The total superfluid fraction is the value at the crust-liquid boundary. 
}
\label{fig-Is_vs_Pressure}
\end{figure}

The baryon densities and corresponding degeneracy pressures in the inner crusts are taken from Ref.~\onlinecite{cha12} and Table I of Ref.~\onlinecite{cha13}.
Results for the ratio, $I_s/I_c$, as a function of $P_{\mbox{\tiny max}}\le P_{\mbox{\tiny crust}}$, are shown in Fig.~\ref{fig-Is_vs_Pressure}.
The superfluid moment of inertia fraction is the value at the crust-liquid boundary. Values of $I_s/I_c$ at $P_{\mbox{\tiny crust}}= 0.389\,\mbox{MeV-fm}^{-3}$ for pure neutron matter, neutrons in the amorphous crust and from Bragg entrainment are $0.880\,,0.782\,,\mbox{and}\,0.285$, respectively.
%
%
\end{document}